\documentclass[preprint,aps ,nofootinbib]{revtex4}
\pdfoutput=1
\usepackage{graphicx,subfigure}%Include figure files
\usepackage{epsfig}
\usepackage{amsmath}
\usepackage{amsfonts}
\usepackage{amssymb}
\usepackage{color}%
\usepackage{dcolumn}% Align table columns on decimal point
\usepackage{slashed}
\providecommand{\U}[1]{\protect\rule{.1in}{.1in}}

\newcommand{\f}{\begin{equation}}
\newcommand{\ff}{\end{equation}}
\newcommand{\fa}{\begin{eqnarray}}
\newcommand{\ffa}{\end{eqnarray}}

\begin{document}
\title{The pseudo-gap phase and the duality in holographic fermionic system with dipole coupling on Q-lattice}
\author{Yi Ling $^{1,3}$}
\email{lingy@ihep.ac.cn}
\author{Peng Liu $^{1}$}
\email{liup51@ihep.ac.cn}
\author{Chao Niu $^{1}$}
\email{niuc@ihep.ac.cn}
\author{Jian-Pin Wu $^{2,3}$}
\email{jianpinwu@mail.bnu.edu.cn}
\affiliation{$^1$ Institute of High Energy Physics, Chinese Academy of Sciences, Beijing 100049, China\\
$^2$Institute of Gravitation and Cosmology, Department of Physics,
School of Mathematics and Physics, Bohai University, Jinzhou 121013, China\\
$^3$ State Key Laboratory of Theoretical Physics, Institute of Theoretical Physics, Chinese Academy of Sciences, Beijing 100190, China}

\begin{abstract}

We classify the different phases by the ``pole-zero mechanism"
for a holographic fermionic system which contains a dipole
coupling with strength $p$ on a Q-lattice background. A
complete phase structure in $p$ space can be depicted in
terms of Fermi liquid, non-Fermi liquid, Mott phase and
pseudo-gap phase. In particular, we find that in
general the region of the pseudo-gap phase in $p$ space is
suppressed when the Q-lattice background is dual to a deep
insulating phase, while for an anisotropic background, we have an
anisotropic region for the pseudo-gap phase in $p$ space as well. In
addition, we find that the duality between zeros and poles always
exists regardless of whether or not the model is isotropic.

{\bf Key Words:} Holographic Q-lattice, ``pole-zero mechanism", Fermionic system

{\bf PACS:} 11.25.Tq, 04.70.Bw
\end{abstract}

\maketitle

\section{Introduction}

Until now, a general theoretical framework for quantum
phases and phase transitions of strongly correlated electron
systems, such as cuprate and other oxides, has not been
established yet. As an alternative approach, AdS/CFT
correspondence~\cite{Maldacena:1997re,Gubser:2002,Witten:1998} is a
powerful tool to attack these strongly correlated problems
and to get some possible clues on the basic principle behind these
numerous correlated electron systems.

Indeed, some exotic states of matter, including Fermi liquid,
non-Fermi liquid, Mott phase and
pseudo-gap phase, have been found or identified by AdS/CFT
correspondence
\cite{SSLee:2008,HongLiu:2009,HongLiu:AdS2,Zaanen:2009,Phillips:PRL,Phillips:PRD,AlsupDuality,Phillips:Duality}.
By adding a probe fermion on RN-AdS background, a non-linear
dispersion relation is observed~\cite{HongLiu:2009},
indicating that a non-Fermi liquid phase emerges. Also, it is
found that the low energy behavior of a fermionic system on
the RN-AdS background is controlled by the $AdS_2$ near horizon
geometry~\cite{HongLiu:AdS2}. Later, the properties of
a fermionic system on Gauss-Bonnet, Lifshitz as well
as hyperscaling violation geometry have been extensively
studied
\cite{1103.3982,1108.6134,1203.0674,1105.1162,JPWu:EDM,Kuang:2014pna,1201.3832,1304.7431,1201.1764,BTZ}.
Furthermore, a dipole coupling between the gauge field and
the Dirac field can be introduced to model the Mott
phase~\cite{Phillips:PRL,Phillips:PRD}. Many extended
works on the dipole coupling effects have been explored in more
general geometries in
\cite{Dipole1,Dipole2,Dipole3,Dipole4,Dipole5,Dipole6,WJLi:Dipole}.
Recently, in terms of the ``pole-zero mechanism", a pseudo-gap
phase has been detected in a holographic fermionic system
with dipole coupling in RN-AdS black hole and the
Schwarzschild-AdS black hole~\cite{AlsupDuality,Phillips:Duality}.
Moreover, they observed a remarkable duality between zeroes
and poles in this holographic system~\cite{AlsupDuality}, which
should be interesting for experimental scientists in
condensed matter physics, although this phenomenon has not been
captured by experiments yet. In this paper, we shall address the
pseudo-gap phase and the duality in a holographic fermionic
system with dipole coupling on Q-lattice geometry.

Motivated by the idea of Q-balls \cite{Coleman:Qball}, a Q-lattice
model in a holographic framework was first constructed in
\cite{Donos:QLattice}, in which the translational symmetry was
broken and a metal-insulator transition observed
through the study of optical conductivity. A lot of relevant
work has been investigated in this context
\cite{QLattice:Generalize,Donos:Hall,Donos:Ther,YiLingQLatticeS,YiLingQLatticeF,YiLingQLatticeHEE,YiLingQLatticeDM}.
In~\cite{YiLingQLatticeF}, we study a holographic fermionic system
with dipole coupling on Q-lattice geometry and find that a Mott
gap opens when the dipole coupling parameter $p$ is beyond a certain
critical value $p_c$. An interesting result is that the Mott gap
opens much more easily when the Q-lattice background is dual to a
deep insulating phase rather than a metallic phase. Here, we
shall probe the pseudo-gap phase in this system by the ``pole-zero
mechanism" and study how the lattice parameters affect the
formation of the pseudo-gap.

Our paper is organized as follows. In Section \ref{QD}, a brief
review of the holographic Q-lattice geometry and the Dirac
equation is presented. Our main results on the classification of
quantum phases of holographic fermionic system with dipole
coupling are presented for isotropic and anisotropic
Q-lattice geometry in Section \ref{IQL} and Section \ref{AIQL},
respectively. Specifically, we will focus on the pseudo-gap
phase and discuss how the lattice parameters affect its formation.
Finally, the conclusion and discussion are presented in
Section \ref{Conclusion}.

\section{Holographic Q-lattice geometry and the Dirac equation}\label{QD}

In this section, we give a brief introduction on the holographic
Q-lattice model which breaks translational symmetry in both
spatial directions and then demonstrate how to simplify the Dirac
equation over a specific Q-lattice background. For detailed
discussion, we refer to
\cite{Donos:QLattice,YiLingQLatticeS,YiLingQLatticeF}.

The action with Q-lattice structure in both of the spatial
directions can be written as
\begin{eqnarray}
&&
\label{Action}
S=\int d^4x\sqrt{-g}\left[R+6-\frac{1}{2}F^2-|\partial\phi_1|^2-m_1^2|\phi_1|^2-|\partial\phi_2|^2-m_2^2|\phi_2|^2\right],
\end{eqnarray}
where $F=dA$. $\phi_1$ and $\phi_2$ are the complex scalar fields
simulating the lattices. From the above action, the
equations of motion can be deduced as
\begin{eqnarray}
&&
\nonumber
R_{\mu\nu}=g_{\mu\nu}(-3+\frac{m_1^2}{2}|\phi_1|^2+\frac{m_2^2}{2}|\phi_2|^2)+(F_{\mu}^{\ \rho}F_{\nu\rho}-\frac{1}{4}g_{\mu\nu}F^2)+\partial_{(\mu}\phi_1\partial_{\nu)}\phi_1^{\ast}+\partial_{(\mu}\phi_2\partial_{\nu)}\phi_2^{\ast},
\\
\label{EOM}
&&
(\nabla^2-m_1^2)\phi_1=0,~~~(\nabla^2-m_2^2)\phi_2=0,~~~\nabla_{\mu}F^{\mu\nu}=0.
\end{eqnarray}
To find solutions to the above equations of motion, we take the
following ansatz
\begin{eqnarray}
&&
\nonumber
ds^2=-g_{tt}(z)dt^2+g_{zz}(z)dz^2+g_{xx}(z)dx^2+g_{yy}(z)dy^2,
\\
\
&&
\phi_1=e^{ik_1x}\varphi_1,~~~~~
\phi_2=e^{ik_2y}\varphi_2,~~~~~
A=A_t(z)dt,
\end{eqnarray}
with
\begin{eqnarray}
&&
\nonumber
g_{tt}(z)=\frac{(1-z)P(z)Q(z)}{z^2},~~~~g_{zz}(z)=\frac{1}{z^2(1-z)P(z)Q(z)},~~
\\
\nonumber
&&
g_{xx}(z)=\frac{V_1(z)}{z^2},~~~~g_{yy}(z)=\frac{V_2(z)}{z^2},~~~~A_t(z)=\mu (1-z)a(z),
\label{Ansatz}
\\
&&
P(z)=1+z+z^2-\frac{\mu^2 z^3}{2}
,
\end{eqnarray}
where $k_1$ and $k_2$ are two wave-numbers along $x$ and $y$ directions, respectively,
{such that the $\phi_{1,2}$ is periodic in $x,y$ direction with lattice constant $2\pi/k_{1,2}$.}
%which relate to the lattice constant $2\pi/k_1$ and $2\pi/k_2$.
In addition, in this paper, we will set $m_{1,2}^2=-5/4$ for definiteness,
which avoids the violation of the $AdS_2$ BF bound near the horizon.

Based on the above ansatz, we obtain five second order ODEs
for $V_1$, $V_2$, $a$, $\varphi_1$, $\varphi_2$ and one first
order ODE for $Q$. To solve the ODEs numerically, we impose a
regular boundary condition at the horizon $z=1$ and impose the
following conditions on the conformal boundary:
\begin{eqnarray}
Q(0)=1,~~V_1(0)=V_2(0)=1,~~a(0)=1.
\end{eqnarray}
We will only focus on the standard quantisation of the scalar
field where the asymptotic behavior of $\varphi_{1,2}$ is
described as $\varphi_{1,2}=\lambda_{1,2} z^{\Delta_{1-,2-}}$,
with the UV behavior corresponding to Q-lattice deformation with lattice amplitude
$\lambda_{1,2}$, where $\Delta_{1\pm,2\pm}=3/2\pm (9/4 + m^2_{1,2})^{1/2}$, is the scaling dimension of the dual field of $\phi_{1,2}$. In
addition, the Hawking temperature of the black hole is given by
\begin{eqnarray}
T=\frac{P(1)Q(1)}{4\pi}.
\end{eqnarray}
As a result, each of our Q-lattice solutions {can be specified} by five
dimensionless quantities $T/\mu$, $\lambda_{1}/\mu^{\Delta_{1-}}$,
$\lambda_{2}/\mu^{\Delta_{2-}}$, $k_{1}/\mu$ and $k_{2}/\mu$.
{We shall abbreviate these quantities to $T,\lambda_{1,2},k_{1,2}$ respectively in what follows for simpleness.}

Now, we introduce the Dirac equation with dipole coupling on this Q-lattice geometry.
We consider the following action, which involves the interaction between the spinor field and the gauge field
\begin{eqnarray}
\label{actionspinor}
S_{D}=i\int d^{4}x \sqrt{-g}\overline{\zeta}\left(\Gamma^{a}\mathcal{D}_{a} - m_{\zeta} - ip\slashed{F}\right)\zeta.
\end{eqnarray}
In the above action,
$\mathcal{D}_{a}=\partial_{a}+\frac{1}{4}(\omega_{\mu\nu})_{a}\Gamma^{\mu\nu}-iq
A_{a}$ and
$\slashed{F}=\frac{1}{2}\Gamma^{\mu\nu}(e_\mu)^a(e_\nu)^bF_{ab}$,
where $(e_{\mu})^{a}$ form a set of
orthogonal normal vector bases and $(\omega_{\mu\nu})_{a}$ is the spin connection 1-forms.
With the redefinition $\zeta = (g_{tt} g_{xx} g_{yy})^{-1/4}\mathcal{F}$, and by denoting the Fourier transform of
$\mathcal F$ as $F(z,{\bf k})\equiv (\mathcal{A}_\alpha,\mathcal{B}_{\alpha})^T$ with $\alpha=1,2$, and choosing the following gamma matrices,
\begin{eqnarray}
\label{GammaMatrices}
 && \Gamma^{3} = \left( \begin{array}{cc}
-\sigma^3  & 0  \\
0 & -\sigma^3
\end{array} \right), \;\;
 \Gamma^{0} = \left( \begin{array}{cc}
 i \sigma^1  & 0  \\
0 & i \sigma^1
\end{array} \right),
\cr
&&
\Gamma^{1} = \left( \begin{array}{cc}
-\sigma^2  & 0  \\
0 & \sigma^2
\end{array} \right), \;\;
 \Gamma^{2} = \left( \begin{array}{cc}
 0  & \sigma^2  \\
\sigma^2 & 0
\end{array} \right).
\end{eqnarray}
the Dirac equation deduced from Eq.(\ref{actionspinor}) can be written in a simple form as
\begin{eqnarray} \label{DiracEAB1}
&&
\left(\frac{1}{\sqrt{g_{zz}}}\partial_{z}\mp m_{\zeta} \right)\left( \begin{matrix} \mathcal{A}_{1} \cr  \mathcal{B}_{1} \end{matrix}\right)
\pm (\omega+ q A_{t})\frac{1}{\sqrt{g_{tt}}}\left( \begin{matrix} \mathcal{B}_{1} \cr  \mathcal{A}_{1} \end{matrix}\right)
+\frac{p}{\sqrt{g_{zz}g_{tt}}}(\partial_{z}A_{t})\left( \begin{matrix} \mathcal{B}_{1} \cr  \mathcal{A}_{1} \end{matrix}\right)
\nonumber
\\
&&
- \frac{k_x}{\sqrt{g_{xx}}} \left( \begin{matrix} \mathcal{B}_{1} \cr  \mathcal{A}_{1} \end{matrix}\right)
+ \frac{k_y}{\sqrt{g_{yy}}} \left( \begin{matrix} \mathcal{B}_{2} \cr  \mathcal{A}_{2} \end{matrix}\right)
=0,
\end{eqnarray}
\begin{eqnarray} \label{DiracEAB2}
&&
\left(\frac{1}{\sqrt{g_{zz}}}\partial_{z}\mp m_{\zeta}\right)\left( \begin{matrix} \mathcal{A}_{2}\cr  \mathcal{B}_{2}\end{matrix}\right)
\pm (\omega+ q A_{t})\frac{1}{\sqrt{g_{tt}}}\left( \begin{matrix} \mathcal{B}_{2}\cr  \mathcal{A}_{2}\end{matrix}\right)
+\frac{p}{\sqrt{g_{zz}g_{tt}}}(\partial_{z}A_{t})\left( \begin{matrix} \mathcal{B}_{2} \cr  \mathcal{A}_{2} \end{matrix}\right)
\nonumber
\\
&&
+\frac{k_x}{\sqrt{g_{xx}}} \left( \begin{matrix} \mathcal{B}_{2}\cr  \mathcal{A}_{2}\end{matrix}\right)
+ \frac{k_y}{\sqrt{g_{yy}}} \left( \begin{matrix} \mathcal{B}_{1}\cr  \mathcal{A}_{1}\end{matrix}\right)
=0.
\end{eqnarray}
To solve the Dirac equation, we need to impose the following independent ingoing boundary condition at the horizon
\begin{equation}
\left( \begin{matrix} \mathcal{A}_{\alpha}(z,\textbf{k}) \cr  \mathcal{B}_{\alpha}(z,\textbf{k}) \end{matrix}\right)
=c_\alpha\left( \begin{matrix} 1
\cr  -i\end{matrix}\right)(1-z)^{-\frac{i\omega}{4\pi T}}.
\end{equation}
{The near boundary behavior of the Dirac field will be}
\begin{equation}
\left( \begin{matrix}
  \mathcal{A}_{\alpha} \cr \mathcal{B}_\alpha
\end{matrix} \right)
\approx
a_\alpha z^{m_\zeta}
\left( \begin{matrix}
  1 \cr 0
\end{matrix} \right)
+
b_\alpha z^{-m_\zeta}
\left( \begin{matrix}
  0 \cr 1
\end{matrix} \right).
\end{equation}
Finally, we can read off the retarded Green function by
holography
\begin{eqnarray}
a_{\alpha}=G_{\alpha\alpha'}b_{\alpha'}.
\end{eqnarray}
To obtain the boundary Green function, we need to construct a
basis of finite solutions,
$(\mathcal{A}^{I}_{\alpha},\mathcal{B}^{I}_{\alpha})$ and
$(\mathcal{A}^{II}_{\alpha},\mathcal{B}^{II}_{\alpha})$ due to the
four components of the Dirac fields being coupled to one another.

\section{Pseudo-gap phases and duality on an isotropic Q-lattice}\label{IQL}

There is competition between the poles ($k=k_F$) and zeros
($k=k_L$) in the Green function of a strongly coupled
fermionic system in condensed matter physics. By the ``pole-zero
mechanism", we can classify the different phases of a strongly
coupled fermionic
system~\cite{Phillips:2007,Dzyaloshinskii:2003,Phillips:2012,Imada:2009,Imada:2010}.
The criterion of the phase classification is given as below.
\[ \begin{aligned}
  \text{Poles} & \quad \Leftrightarrow \quad \text{(Non-)Fermi liquid phase}\\
  \text{Zeros} & \quad \Leftrightarrow \quad \text{Mott Insulator phase}\\
    \text{Coexistence of poles and zeros} & \quad \Leftrightarrow \quad \text{Pseudo-gap phase}
\end{aligned} \]
In a holographic framework, this ``pole-zero mechanism" was first
introduced in \cite{AlsupDuality} to characterize different
phases. Here, we will use this mechanism to probe the pseudo-gap
phase in the holographic fermionic system with dipole coupling on
Q-lattice geometry. We explore the case of an isotropic Q-lattice in
this section, i.e., $\lambda_1=\lambda_2$ and $k_1=k_2$,
in which we can set $k_y=0$ without loss of generality.
The case of anisotropic Q-lattice geometry will be discussed in the next section.

For definiteness, we fix $m=0$ as well as $q=1$ and
work at an extremely low temperature $T\simeq 0.00398$ throughout this paper.

\begin{figure}
\centering
\subfigure[]{\label{p0:a}
\includegraphics[scale=0.6]{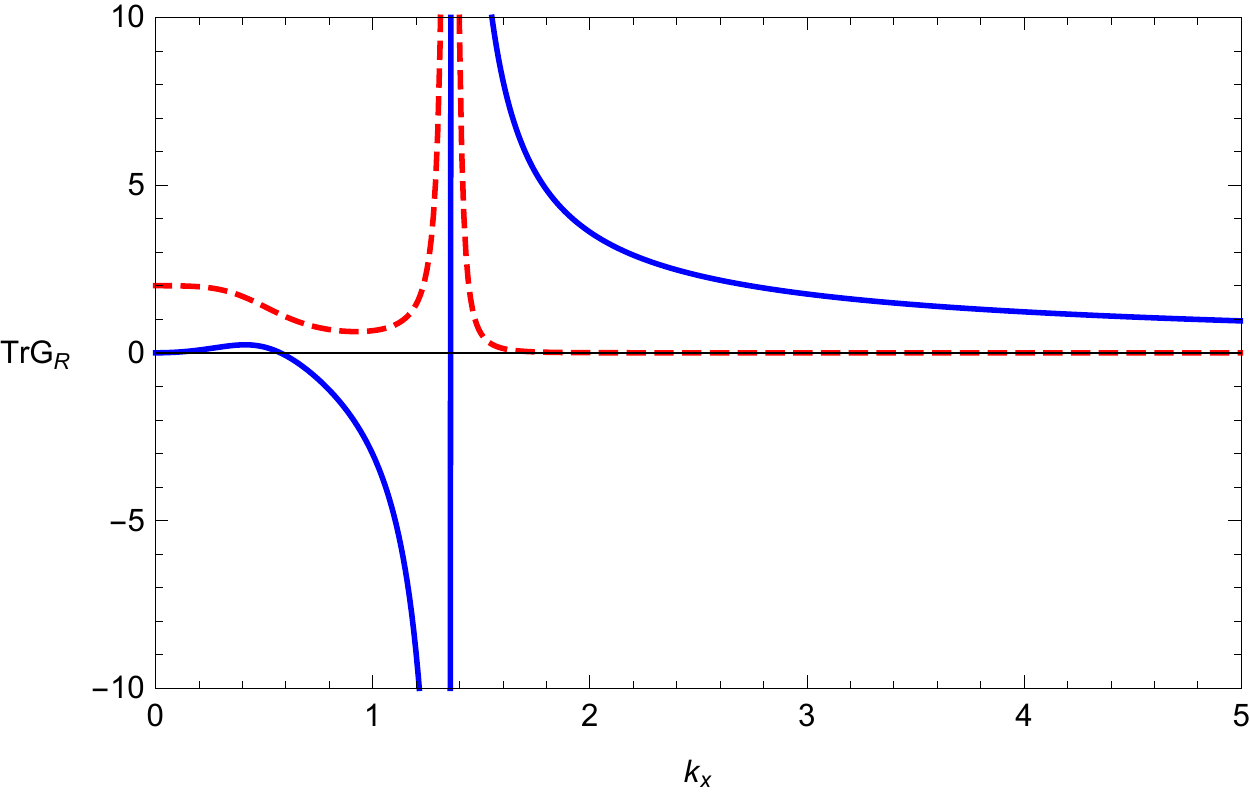}\label{tua}\hspace{0.2cm}}
\subfigure[]{\label{p0:b}
\includegraphics[scale=0.6]{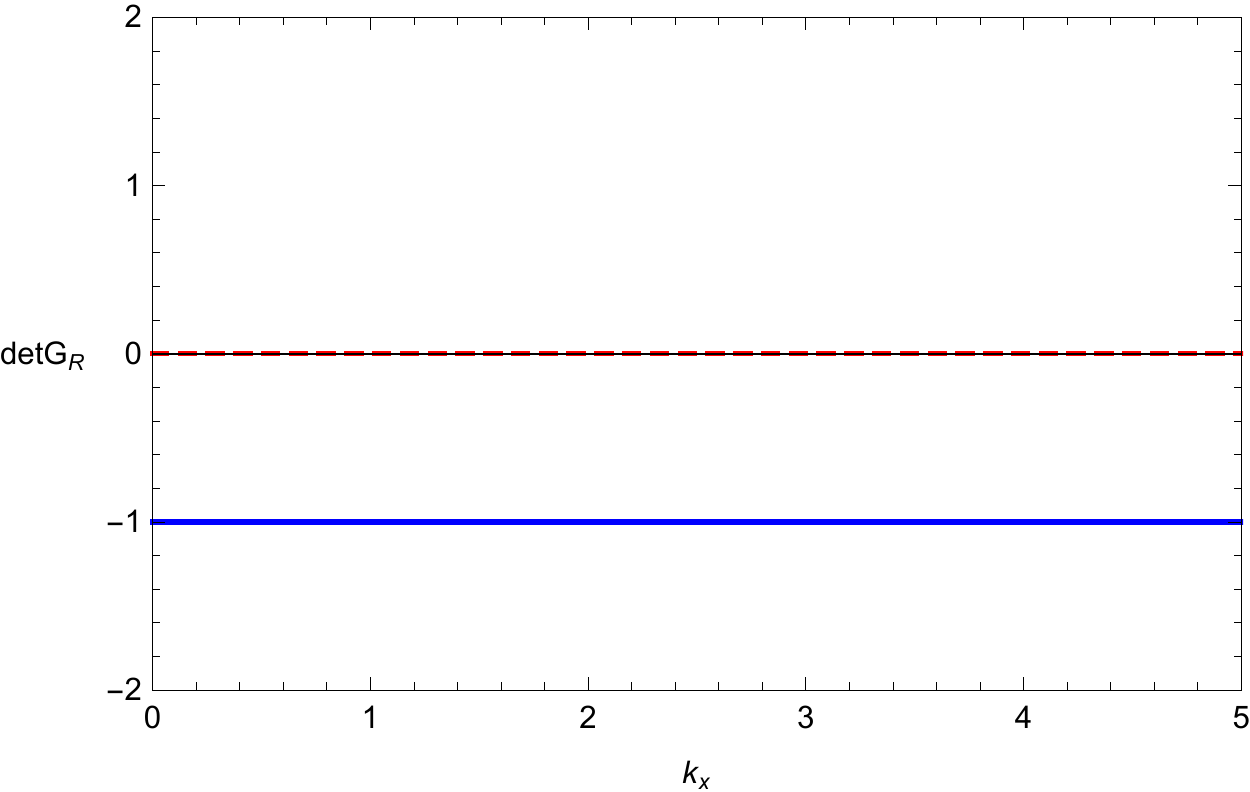}\label{tub}\hspace{0.2cm}}
\caption{\label{p0} (a): $\text{Re(Tr} G_R)$ (solid blue line) and $\text{Im(Tr} G_R)$ (dashed red line) for $p=0$ at $\omega=0$.
A pole is visible in the spectral function $A(k_x)$ ($\text{Im(Tr} G_R)$), which indicates a Fermi surface ($k_F\simeq 1.359$).
(b): $\text{Re}(det G_R)$ (solid blue line) and $\text{Im(} det G_R)$ (dashed red line) for $p=0$ at $\omega=0$,
in which neither poles nor zeroes is observed.
Here, we have fixed $\lambda_1=\lambda_2=0.5$ and $k_1=k_2=0.8$.
}
\end{figure}

\begin{figure}
\center{
\subfigure[]{\label{p4p5:a}
\includegraphics[scale=0.6]{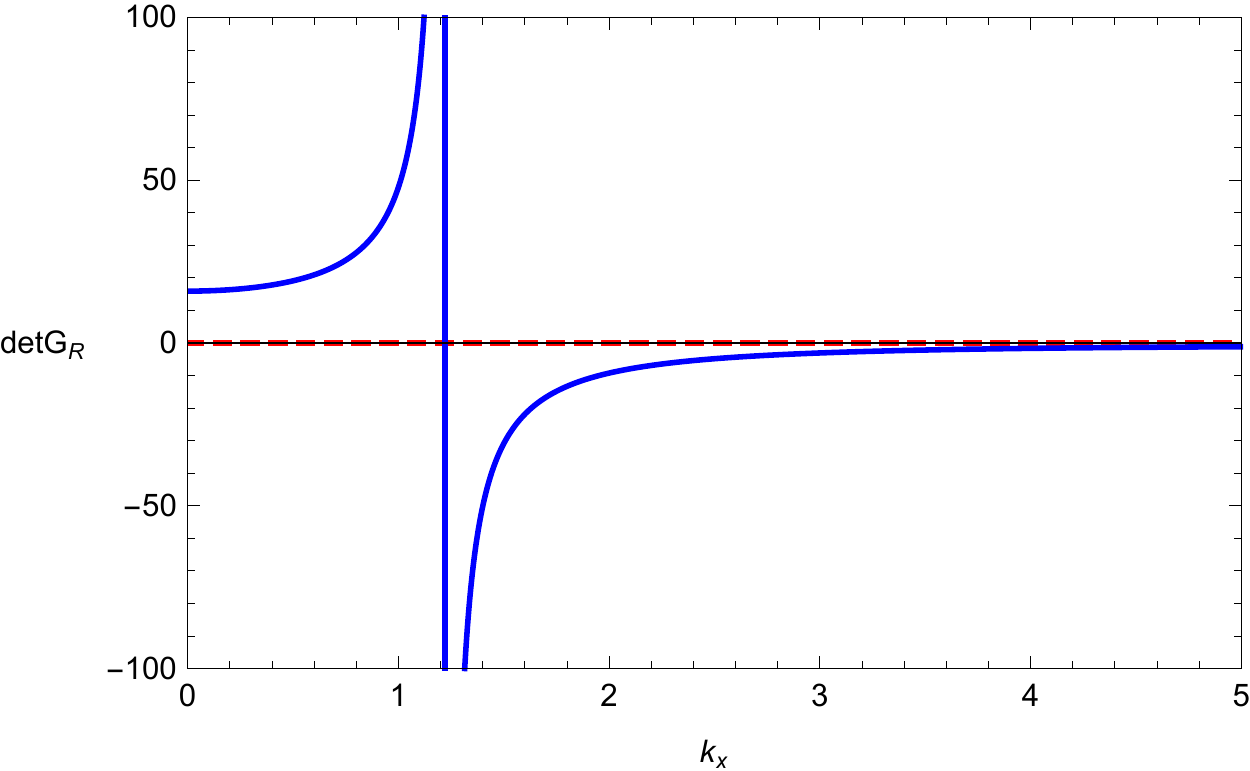}\hspace{0.2cm}}
\subfigure[]{\label{p4p5:b}
\includegraphics[scale=0.6]{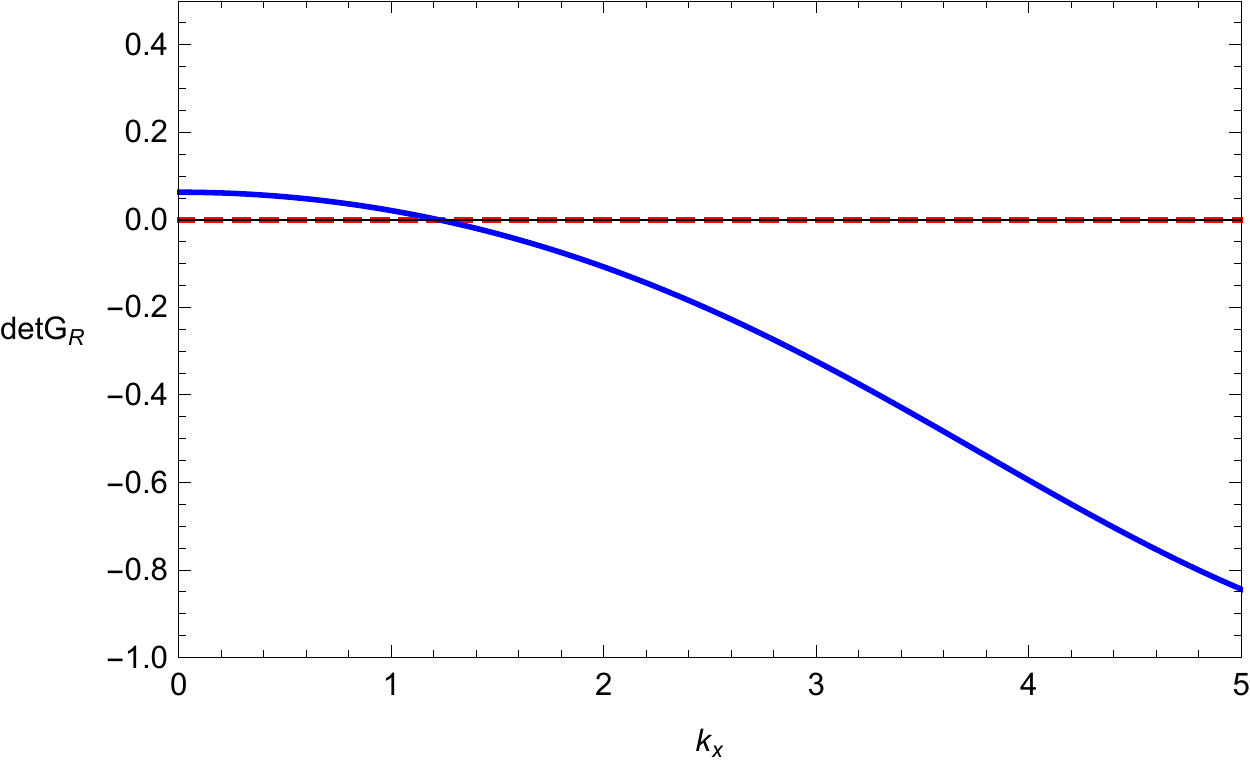}\hspace{0.2cm}}
\caption{\label{p4p5}$\text{Re} (det G_R)$ (solid blue line) and $\text{Im} (det G_R)$ (dashed red line) at $\omega=0$.
  (a) ($p=-4.5$) shows a pole at $k_x=k_F\simeq 1.222$ and (b) ($p=4.5$) shows a zero at $k_x=k_L\simeq1.222$.
Here, we have fixed $\lambda_1=\lambda_2=0.5$ and $k_1=k_2=0.8$.}}
\end{figure}

\begin{figure}
\center{
\subfigure[]{\label{p0p1:a}
\includegraphics[scale=0.6]{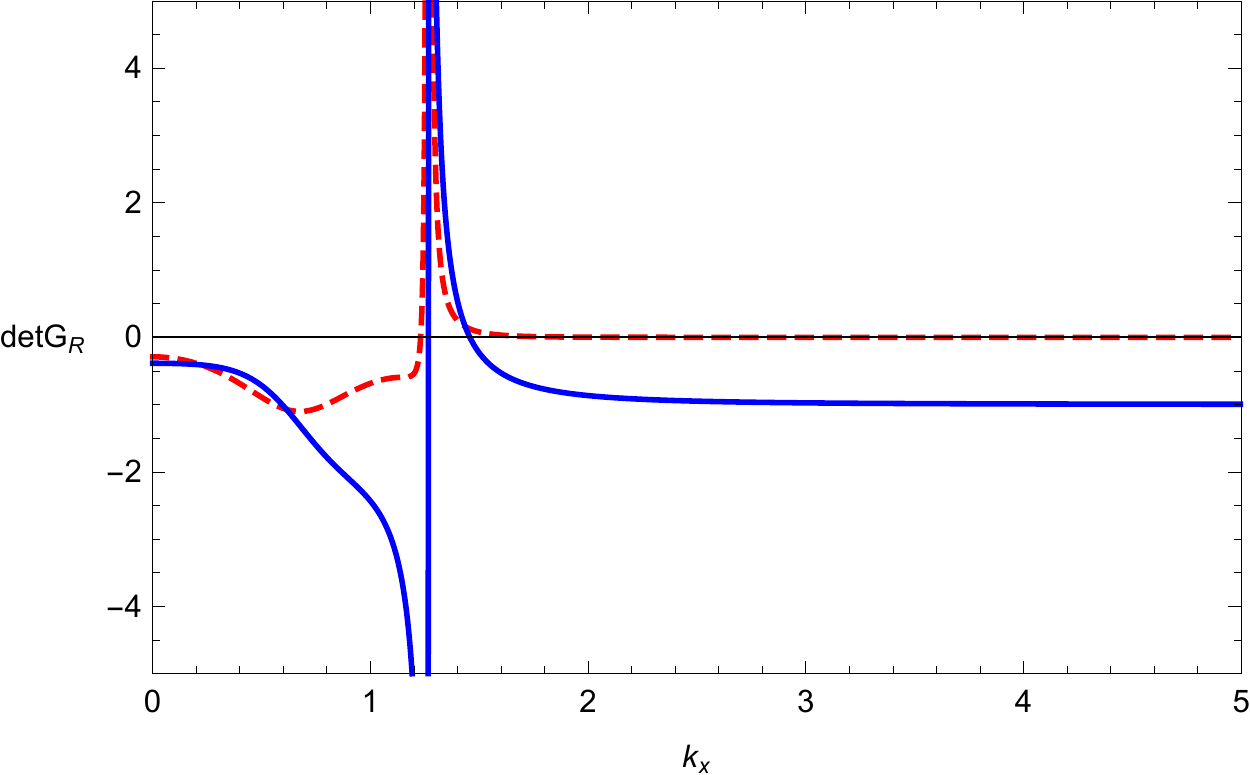}\hspace{0.2cm}}
\subfigure[]{\label{p0p1:b}
\includegraphics[scale=0.6]{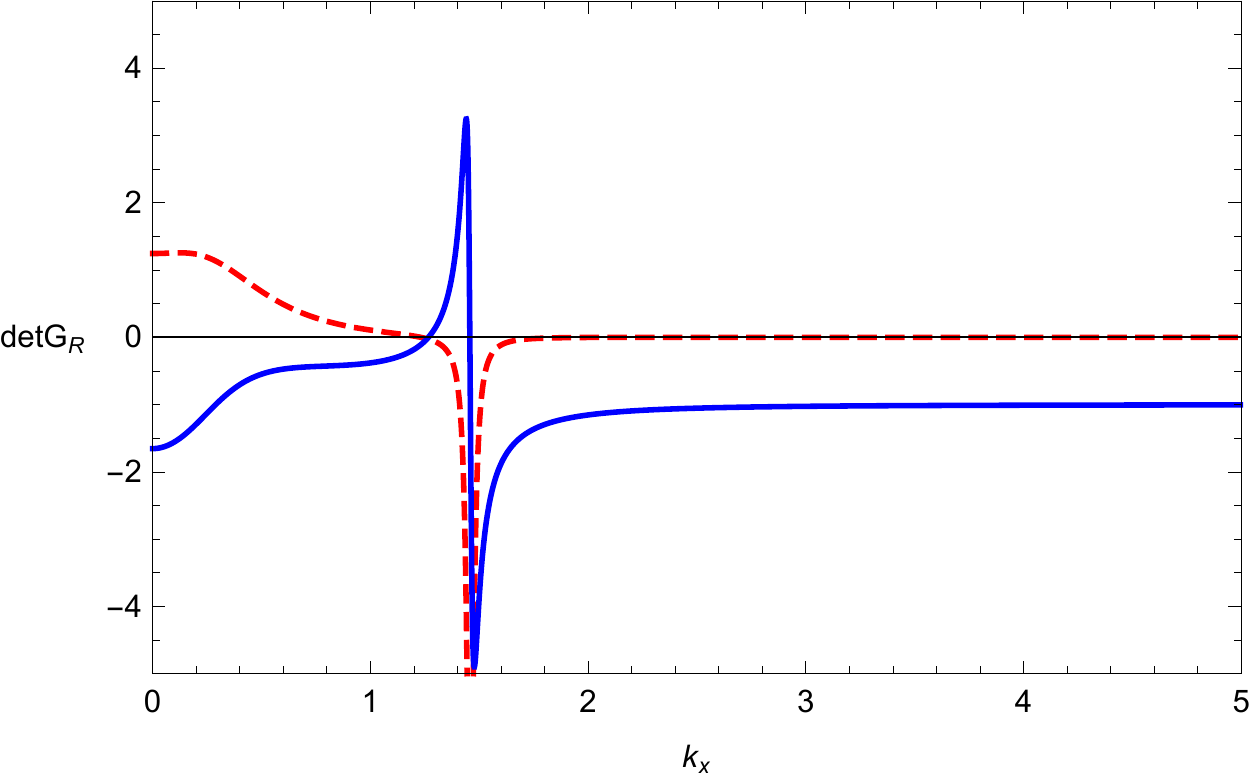}\hspace{0.2cm}}
\caption{\label{p0p1}$\text{Re} (det G_R)$ (solid blue line)
and $\text{Im} (det G_R)$ (dashed red line) at $\omega=0$.
Both pole and zero can be seen for $p=-0.1$ ((a):
$k_x=k_L\simeq 1.441$ and $k_x=k_F\simeq 1.274$) and $p=0.1$ ((b): $k_x=k_L\simeq 1.274$ and $k_x=k_F\simeq 1.441$). Here, we have
fixed $\lambda_1=\lambda_2=0.5$ and $k_1=k_2=0.8$.}}
\end{figure}

Firstly, for $p=0$, neither poles nor zeros can be observed in
the determinant of the Green function $det G_R$ because the
poles (zeros) of Green function are cancelled
(Fig. \ref{p0}). However, once the dipole coupling is turned on,
the poles or zeros emerge in $det G_R$ so that we can classify the
phases for the fermionic system with dipole coupling in terms of
the ``pole-zero mechanism". Fig. \ref{p4p5} shows the determinant of
the Green function $det G_R$ as a function of $k_x$ at $\omega=0$ for
$p=-4.5$ (Fig. \ref{p0:a}) and $p=4.5$ (Fig. \ref{p0:b}) with $\lambda_1=\lambda_2=0.5$
and $k_1=k_2=0.8$. One pole emerges at $k_x=k_F\simeq 1.222$ for $p=-4.5$,
indicating a (non-)Fermi liquid state, and a zero can be found at
the same momentum $k_x=k_L\simeq1.222$ for $p=4.5$, indicating a
Mott state. Obviously, there is a duality between zero and pole
under $p\rightarrow -p$, which was first revealed in
\cite{AlsupDuality}. When we decrease the dipole coupling to
$|p|=0.1$, we find the coexistence of pole and zero in $det
G_R$ (Fig. \ref{p0p1}), which points to a pseudo-gap phase.
Again, the duality between zeros and poles remains under
$p\rightarrow -p$ (Fig. \ref{p0p1}). According to the above
observations, we claim that the (non-)Fermi liquid phase, Mott phase
and the pseudo-gap phase emerge in a fermionic system with dipole
coupling on the Q-lattice. By the density of state (DOS) $A(\omega)$,
we can determine the critical point $p_c$ of the transition from
(non-)Fermi liquid phase to Mott phase and find that the Mott gap
opens more easily in a deep insulating phase than a metallic phase
\cite{YiLingQLatticeF}. Now, we shall focus on the effect of
lattice constant $k_{1,2}$ and lattice amplitude $\lambda_{1,2}$ on the
pseudo-gap phase. By careful numerical calculation, we find
that a pseudo-gap emerges when $|p|\lesssim 0.605$ for
$\lambda_1=\lambda_2=0.5$ and $k_1=k_2=0.8$. When we go to a deep insulating phase
with $\lambda_1=\lambda_2=2$ and $k_1=k_2=1/2^{3/2}$, the pseudo-gap phase emerges
in the region of $|p|\lesssim 0.335$. Therefore, the region
of the pseudo-gap phase in $p$ space is suppressed in the deep
insulating phase. For comparison, we also find that the pseudo-gap
phase appears when $|p|\lesssim 0.634$ in the RN-AdS black
hole background, which is obtained simply by setting $\lambda_1=\lambda_2=0$
\footnote{Because we have set the gauge coupling
constant $g_F=\sqrt{2}$ here, which is different from the
conventions in \cite{AlsupDuality} that $g_F=2$, the charge $q$
and dipole coupling $p$ will correspond to $\sqrt{2}q$ and
$\sqrt{2}p$ in \cite{AlsupDuality} due to the products $g_Fq$ and
$g_Fp$ being the relevant quantities. }.
%%%%%%%%%%%%%%%%%%%%%%%%%%%%%%
%%%%%%%%%%%%%%%%%%%%%%%%%%%%%%
%%%%%%%%%%%%%%%%%%%%%%%%%%%%%%
%%%%%%%%%%%%%%%%%%%%%%%%%%%%%%

%%%%%%%%%%%%%%%%%%%%%%%%%%%%%%
%%%%%%%%%%%%%%%%%%%%%%%%%%%%%%
%%%%%%%%%%%%%%%%%%%%%%%%%%%%%%
%%%%%%%%%%%%%%%%%%%%%%%%%%%%%%

%%%%%%%%%%%%%%%%%%%%%%%%%%%%%%
%%%%%%%%%%%%%%%%%%%%%%%%%%%%%%

Before closing this section, we would like to present a brief discussion on the duality between zeros and poles.
When $k_y$ is set to zero, we can package the Dirac equations (\ref{DiracEAB1}) and (\ref{DiracEAB2}) into the following evolution equations
\begin{eqnarray} \label{DiracEF1}
(\partial_{z}-2m_{\zeta}\sqrt{g_{zz}}) \xi_{\alpha}
+\left[ v_{-} + (-1)^{\alpha} k \sqrt{\frac{g_{zz}}{g_{xx}}}  \right]
+ \left[ v_{+} - (-1)^{\alpha} k \sqrt{\frac{g_{zz}}{g_{xx}}}  \right]\xi_{\alpha}^{2}
=0
~,
\end{eqnarray}
where we have defined $\xi_{\alpha}\equiv \frac{\mathcal{A}_{\alpha}}{\mathcal{B}_{\alpha}}$ and $v_{\pm}=\sqrt{\frac{g_{zz}}{g_{tt}}}(\omega+q A_{t})\pm p\sqrt{g^{tt}} \partial_{z} A_{t}$.
For massless fermions, the retarder Green function is
\begin{eqnarray} \label{GreenFBoundary}
G (\omega,k)= \lim_{z\rightarrow 0}
\left( \begin{array}{cc}
\xi_{1}   & 0  \\
0  & \xi_{2} \end{array} \right)  \ .
\end{eqnarray}
Therefore, from the evolution equation (\ref{DiracEF1}), one can easily find the following symmetry of Green's function
\begin{eqnarray} \label{sym1}
G_{11}(\omega,k)=G_{22}(\omega,-k)\ .
\end{eqnarray}
In addition, we can introduce the reciprocal of $\xi_{\alpha}$,
\begin{eqnarray}
\slashed{\xi}_{\alpha}=\frac{1}{\xi_{\alpha}}
~.
\end{eqnarray}
We find that $\slashed{\xi}_{\alpha}$ satisfies the following evolution equations
\begin{eqnarray} \label{DiracEF2}
(\partial_{z}-2m_{\zeta}\sqrt{g_{zz}}) \slashed{\xi}_{\alpha}
+\left[ \slashed{v}_{-} + (-1)^{\alpha} k \sqrt{\frac{g_{zz}}{g_{xx}}}  \right]
+ \left[ \slashed{v}_{+} - (-1)^{\alpha} k \sqrt{\frac{g_{zz}}{g_{xx}}}  \right]\slashed{\xi}_{\alpha}^{2}
=0
~,
\end{eqnarray}
where $\slashed{v}_{\pm}=-\sqrt{\frac{g_{zz}}{g_{tt}}}(\omega+q A_{t})\pm p\sqrt{g^{tt}} \partial_{z} A_{t}$.
It is easily found that there is a symmetry between Eqs. (\ref{DiracEF1}) and (\ref{DiracEF2}) under the transformation $k\rightarrow -k$,
$p\rightarrow -p$ and $\xi_{\alpha}\rightarrow -\slashed{\xi}_{\alpha}$.
Combining Eq. (\ref{sym1}), we can easily find that there is a duality between zeros and poles when $p\rightarrow -p$.

\section{Pseudo-gap phases and duality on an anisotropic Q-lattice}\label{AIQL}

Now we turn to study the pseudo-gap phase and the duality
between the zeros and poles on
anisotropic Q-lattices, in which we set $\lambda_1=1$,
$\lambda_2=0.1$ and $k_1=k_2=0.8$.

\begin{figure}
\center{
\subfigure[]{\label{kxp6:a}
\includegraphics[scale=0.6]{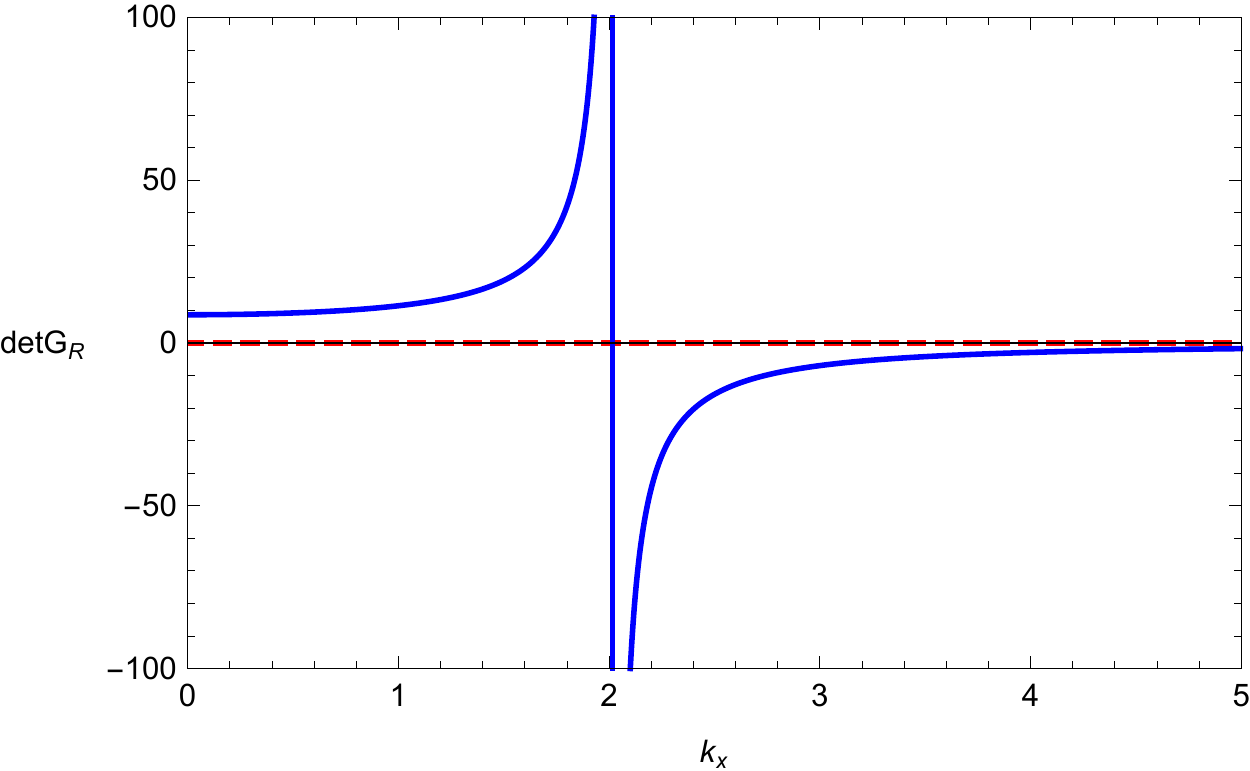}\hspace{0.2cm}}
\subfigure[]{\label{kxp6:b}
\includegraphics[scale=0.6]{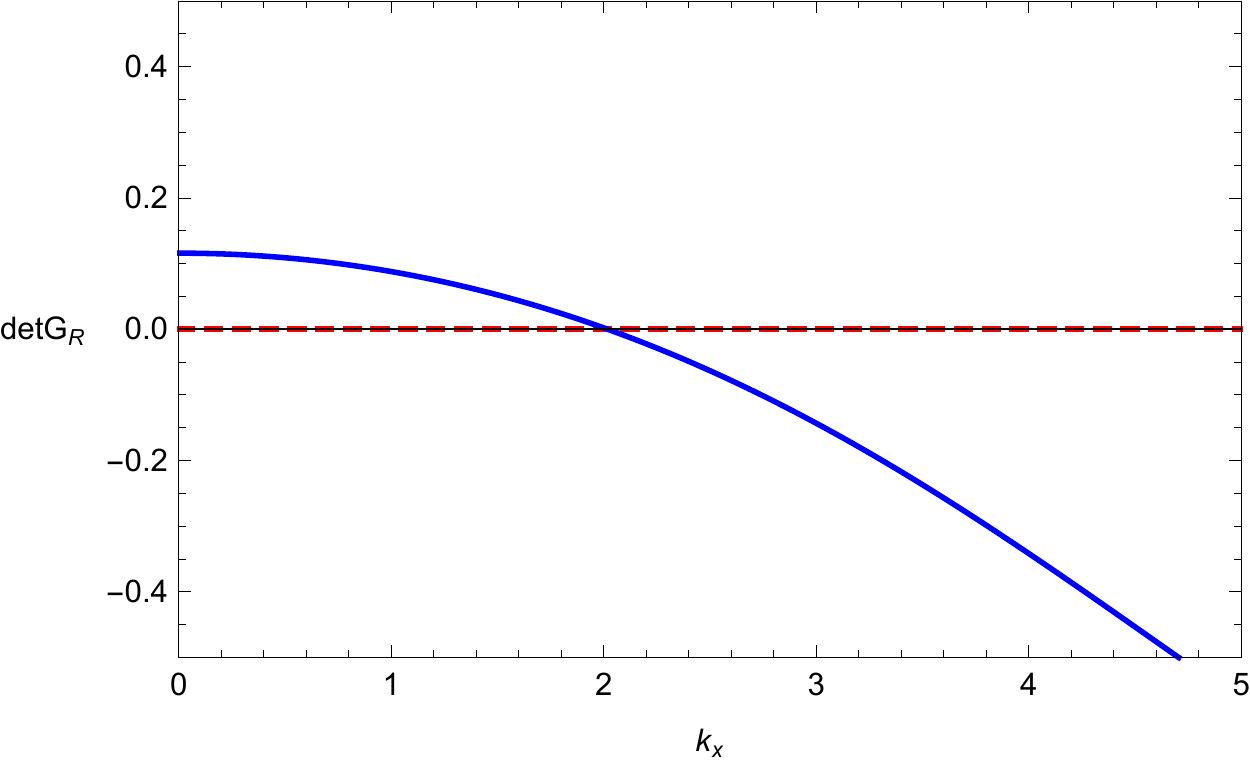}\hspace{0.2cm}}
\caption{\label{kxp6}$\text{Re} (det G_R)$ (solid blue line) and $\text{Im} (det G_R)$ (dashed red line) as a function of $k_x$ (we set $k_y=0$)
at $\omega=0$ with $p=-6$ ((a)) and $p=6$ ((b))
on anisotropic Q-lattice background with $\lambda_1=1$, $\lambda_2=0.1$ and $k_1=k_2=0.8$.
 (a) shows a pole at $k_x = k_{xF}\simeq 2.016$ and (b) shows a zero at $k_x=k_{xL}\simeq2.016$.}}
\end{figure}

\begin{figure}
\center{
\subfigure[]{\label{kxp0p1:a}
\includegraphics[scale=0.6]{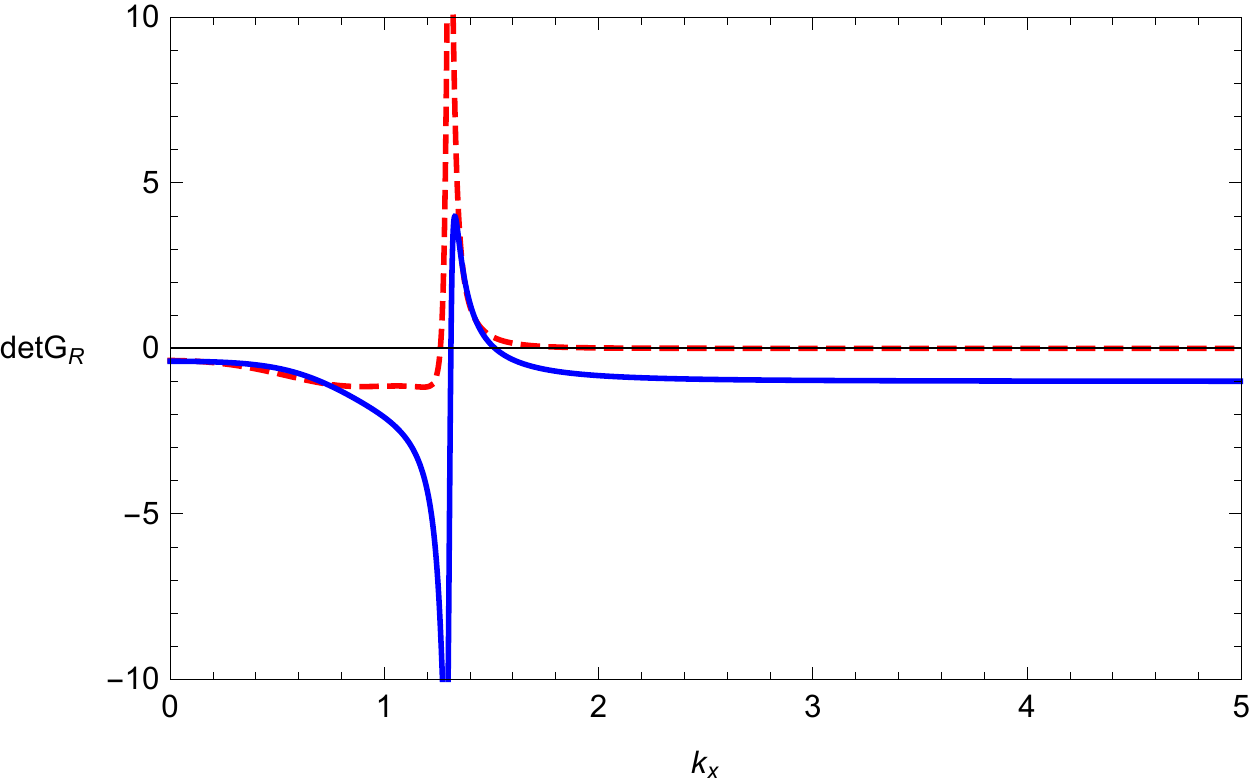}\hspace{0.2cm}}
\subfigure[]{\label{kxp0p1:b}
\includegraphics[scale=0.6]{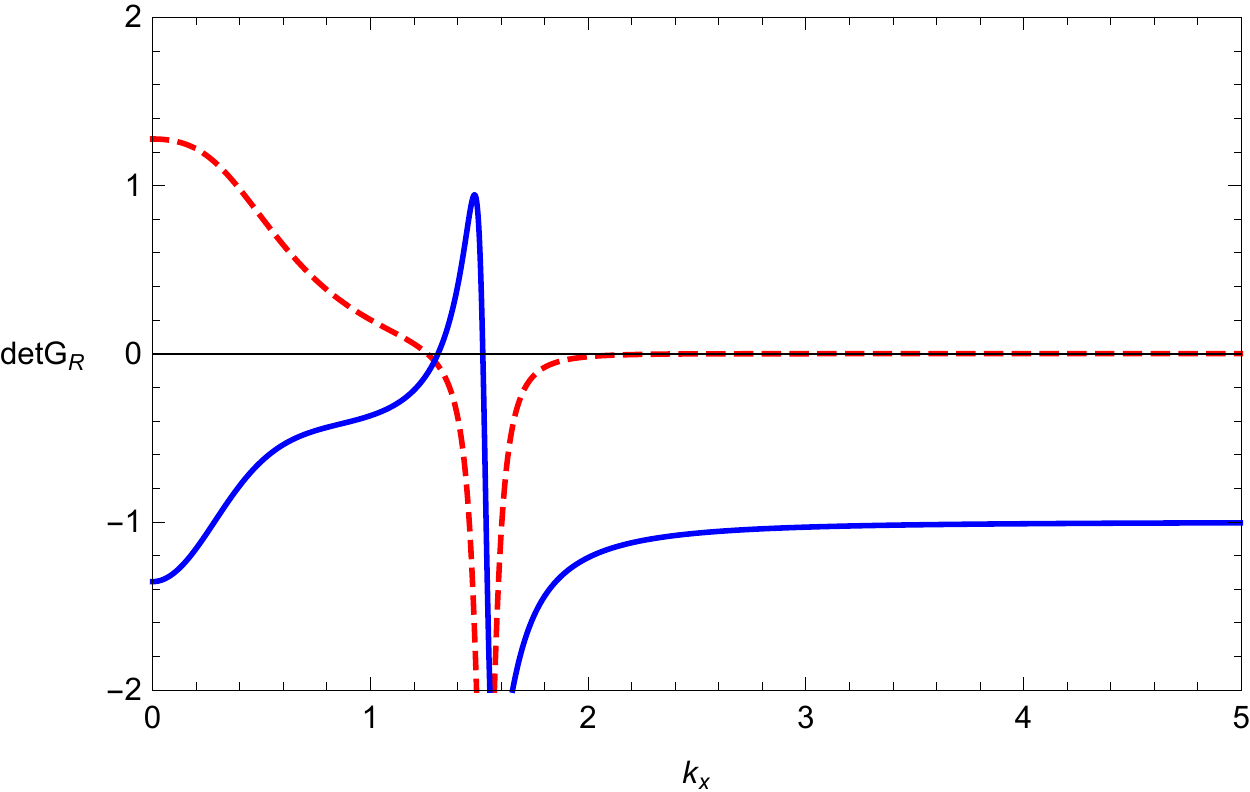}\hspace{0.2cm}}
\caption{\label{kxp0p1}$\text{Re} (det G_R)$ (solid blue line)
and $\text{Im} (det G_R)$ (dashed red line) as a function of
$k_x$ (we set $k_y=0$) at $\omega=0$ with $p=-0.1$ ((a)) and
$p=0.1$ ((b)) on anisotropic Q-lattice background with
$\lambda_1=1$, $\lambda_2=0.1$ and $k_1=k_2=0.8$.. Both pole and
zero can be seen for $p=-0.1$ ($k_x=k_{xL}\simeq 1.479$ and
$k_x=k_{xF}\simeq 1.329$) and $p=0.1$ ($k_x=k_{xL}\simeq 1.329$
and $k_x=k_{xF}\simeq 1.479$). }}
\end{figure}

Firstly, we explore the duality along the $k_x$ direction (setting
$k_y=0$). As illustrated in Fig. \ref{kxp6} and
Fig. \ref{kxp0p1}, we find that the (non-)Fermi liquid phase, Mott
insulating phase and the pseudo-gap phase along the $k_x$ direction
emerge, depending on the values of the dipole coupling parameter
$p$. Furthermore, the duality between zeros and poles still
holds, which can also be understood by a similar analysis to the last section.
In a parallel manner, we also work out the location
of poles and zeros in $det G_R$ along the $k_y$ direction (setting
$k_x=0$) for $p=6$ and $p=-6$. We find that $k_{yF}\simeq 1.983$
for $p=-6$ and $k_{yL}\simeq 1.983$ for $p=6$, which indicates
that the duality remains valid in the $y$ direction. But
obviously we notice that $k_{xF}\neq k_{yF}$ ($k_{xL}\neq
k_{yL}$), reflecting the anisotropy of the system.

We now study the pseudo-gap phase. We find that the pseudo-gap
phase emerges along the $k_x$ direction for $|p|\lesssim 0.393$,
while along the $k_y$ direction for $|p| \lesssim 0.860$. Obviously,
the anisotropic geometry results in the anisotropic pseudo-gap
phase region. Along the insulating direction ($k_x$ direction),
the region of pseudo-gap phase is suppressed, which is consistent
with that explored on isotropic Q-lattices in the previous
section.

%%%%%%%%%%%%%%%%%%%%%%%%%%%%%%
%%%%%%%%%%%%%%%%%%%%%%%%%%%%%%

%%%%%%%%%%%%%%%%%%%%%%%%%%%%%%
%%%%%%%%%%%%%%%%%%%%%%%%%%%%%%

\section{Conclusion and discussion}\label{Conclusion}

By the ``pole-zero mechanism" we have classified the phases
which appear in a holographic fermionic system with dipole coupling
on Q-lattice geometry. These phases could be Fermi liquid,
non-Fermi liquid, Mott phase or pseudo-gap phase, depending on
the strength of the dipole coupling. Therefore, varying the dipole
coupling parameter $p$ can induce quantum phase transitions in our
holographic system. By investigating the spectral function, we can
work out the critical value $p_c$ of the phase transition from
(non-)Fermi liquid to Mott phase and see how the lattice
parameters $\lambda_{1,2}$ and $k_{1,2}$ affect the formation of the Mott phase in
\cite{YiLingQLatticeF}. Here, we have further computed the
determinant of the retarded Green function and focused on the
formation of the pseudo-gap phase. We find that the region of the
pseudo-gap phase is suppressed in the deep insulating phase. For
the anisotropic Q-lattice geometry, we have obtained
an anisotropic pseudo-gap phase region which is also suppressed
along the insulating direction. Another interesting result is
that the duality between zeros and poles previously found
in \cite{AlsupDuality} still holds in either isotropic or
anisotropic Q-lattice geometry and is independent of the lattice
parameters.

\begin{acknowledgments}

We are grateful to Xiao-Mei Kuang for valuable discussions.
This work is supported by the Natural Science Foundation of China
under Grant Nos.11275208, 11305018 and 11178002. Y.L. also
acknowledges the support from Jiangxi young scientists (JingGang
Star) program and 555 talent project of Jiangxi Province. J. P. Wu
is also supported by Program for Liaoning Excellent Talents in
University (No. LJQ2014123).

\end{acknowledgments}

\end{document}